\documentclass[onecolumn,floats]{article}
\usepackage{authblk}
\usepackage{epsfig}
\usepackage{amsmath}
\usepackage{graphicx}
\usepackage{color}

\title{ A Primer on Weyl Semimetals: Down the Discovery of Topological Phases}
\author[1]{Satyaki Kar}
\author[2]{Arun M. Jayannavar}
\affil[1]{AKPC Mahavidyalaya, Bengai, West Bengal 712611, India.}
\affil[2]{Institute of Physics, Bhubaneswar, Odisha 751005, India.}

\begin{document}



    \maketitle
\begin{abstract}
  {\bf Recently discovered Weyl semimetals (WSM) have found special place in topological condensed matter studies for they represent first example of massless Weyl fermions found in condensed matter systems. A WSM shows gapless bulk energy spectra with Dirac-like point degeneracies, famously called Weyl nodes, which carry with themselves well defined chiralities and topologically protected chiral charges. One finds the Berry curvature of the Bloch bands to become singular, like in a magnetic monopole, at these Weyl nodes. Moreover, these systems feature topological surface states in the form of open Fermi arcs. In this review, we undergo a concise journey from graphene based Dirac physics to Weyl semimetals: the underlying Hamiltonians, their basic features and their unique response to external electric and magnetic fields in order to provide a basic walk-through of how the Weyl physics unfolded with time starting from the discovery of Graphene.}
\end{abstract}
{\bf keywords:} Dirac equation; Weyl fermion; Magnetic monopole; Fermi arc; Chiral anomaly.

\section*{Introduction}

In this short review, we  give a chronological account of how the Weyl physics emerges in the field of condensed matter and the interesting nontrivial topologies associated with them. As this descends right from the Dirac theory of fermions, it will be appropriate to introduce them briefly so that the physics discussed become much more clearer as we go on unraveling the basics of the Weyl semimetals.

In quantum mechanics, an wave function (given as $\psi({\bf r},t)$) describes the quantum state of the system/particle. It contains probabilistic information of the position ${\bf r}$ of a particle at time $t$. The time dependence of this function lead us to the study of the Schrodinger equation which shows the dynamics of this $\psi$ as, 
\begin{equation}
 \boxed{i\hbar\frac{\partial\psi}{\partial t}={\hat H}\psi}
\end{equation}
Here $\hat H$ denotes a Hermitian operator called Hamiltonian and it describes operation on the quantum state to obtain the energy of the state. Saying more precisely, if $\psi$ is an eigen-function for $\hat H$, we get $\hat H\psi=E\psi$, where $E$ denotes the energy eigenvalue.
In non-relativistic quantum mechanics, this energy is just the sum of kinetic and potential energies and we can form an operator equation
\begin{equation}
 \hat H=\frac{{\hat p}^2}{2m}+\hat V~~~{\rm or~energy~eigenvalue}~~E=\frac{p^2}{2m}+V
\end{equation}
 where $\hat p$, $\hat V$ are the momentum and potential energy operators while $p$ and $V$ are the corresponding eigenvalues. Here $m$ denote mass of the particle. So it says that energy has quadratic dependence to momentum $p$ (considering no $p$ dependence in $V$).
 
 However, relativistically the energy momentum relation looks like 
\begin{equation}
 E^2=p^2c^2+m^2c^4
\end{equation}
where $m$ and $c$ are rest mass of the body and speed of light in vacuum respectively. So this equation relates square of energy to the square of momentum and 
demonstrates a linear dependence between $E^2$ and $p^2$.
 In order to obtain the expression for the Hamiltonian in such cases, Dirac proposed an Hamiltonian to be
\begin{equation}
\hat H=c{\bf p.\alpha}+mc^2\beta.
\end{equation}
Here $\alpha$ and $\beta$ are matrices with dimensionality depending on the spatial dimensionality of the system and they satisfy Clifford algebra namely, $\{\alpha_i,\alpha_j\}=2\delta_{ij}$, $\{\alpha_i,\beta\}=0$ and $\beta^2=1$.
Combining this with the Schrodinger equation (and remembering that $p_j=-i\hbar{\partial_j}$), we get
\begin{equation}
 i\hbar\frac{\partial\psi}{\partial t}=(-i\hbar c\alpha_i\partial_i+mc^2\beta)\psi.
 \label{eq5}
\end{equation}
Relativity tells us that space and time are connected and together we can form four-vectors with one temporal component (denoted by index 0) and three spatial components (denoted by indices 1,2,3).
Here we define two 4-vectors: $\gamma^\mu=(\gamma^0,\gamma^1,\gamma^2,\gamma^3)$
and $\partial_\mu=(\partial_0,\partial_1,\partial_2,\partial_3)$ where $\partial_0=\frac{\partial}{c\partial t}$ and $\partial_i=\frac{\partial}{\partial x_i}$. With $\gamma^0=\beta$ and $\gamma^i=\beta\alpha_i$ (for $i=1,2,3$) we can rewrite Eq.\ref{eq5} as
\begin{equation}
 \boxed{(i\hbar\gamma^\mu\partial_\mu-mc)\psi=0}
 \label{dirac}
\end{equation}
with $\mu~\rightarrow~0,1,2,3$ denote the running indices for four components 
($i.e.,$ Eq.\ref{dirac} implies a sum over all four $\mu$ indices).\\
\fbox{%
\parbox{1.\linewidth}{%
Upper~and~lower~indices~in~$\gamma^\mu$~and~ $\partial_\mu$~ signify~contravariant~ and~ covariant\cite{tensor} nature~of~ the~respective~4-vectors.}%
}
\\
This is the famous Dirac equation describing the dynamics of Dirac fermions. We should mention here that the Dirac spinor $\psi$ is a complex function in general, with each Dirac fermion consisting of two real components, called Majorana fermions. Moreover, one can get a Weyl fermion when the mass term in the equation vanishes ($i.e.,~m=0$). At the same time, for being a Weyl fermion it also need to have definite chirality (a property we will discuss later) that makes Weyl fermions available only in odd spatial dimensions. 

Though originally developed for relativistic high energy physics, Dirac fermions got its firm existence in low-energy condensed matter systems like graphene, topological insulators etc. In fact, massless Weyl fermions are first observed in condensed-matter systems
called Weyl semimetals and that is what we are going to discuss in this review.

\subsection*{Graphene Dirac fermions}
Dirac physics in condensed matter systems ushered in renewed interests among
the physics community since the famous discovery of exfoliation technique of graphene monolayers in 2004\cite{graphene}.
It showed how a single two-dimensional (2D) sheet scratched off a non-conducting three dimensional (3D) graphite lump can show unique conducting properties such as large electron mobility, thermal conductivity or huge tensile strengths.
Dispersion near the band crossings of such monolayers are linear and results in Dirac fermions for the low energy excitations\cite{neto}.
A simplified Hamiltonian of such system can be constructed from a tight-binding model (by which electron in an orbital localized around a lattice site can move/tunnel to a different orbital localized around an adjacent lattice site) consisting of nearest-neighbor hopping of electrons in the underlying honeycomb lattice. It produces the so-called Dirac points (DP) where the valence and conduction bands touch, the dispersion being linear at that point.
Due to this point degeneracy between conduction and valence band, graphene is 
dubbed as a semimetal.
There are two such independent touching points called Dirac points (corresponding to two separate wave-vectors, K and K',say) within the Brillouin zone (BZ) of the 2D lattice. \\
\fbox{%
\parbox{1.\linewidth}{%
In a periodic crystal lattice, we can find an unit cell, $i.e.,$ a minimum volume of space that can be translated via lattice vectors to 
cover the entire lattice space without any overlap. Like the periodic lattice in real space, there is also a momentum space or $k$-space which constitutes a reciprocal space
corresponding to the lattice. This $k$-space consists of periodically arranged $k$-points which are wave-vectors of the plane waves having same periodicity as the lattice. 
There is an unit cell in this reciprocal space as well and this is called the Brillouin zone\cite{kittel}.}%
}
\\Considering low energy physics around these points, one comes up with
a continuum model that resembles a massless Dirac Hamiltonian.
Typically, a continuum model for graphene is given as $H=\hbar v_F(\sigma_xk_x+\sigma_yk_y)$. Here $v_F$ is the Fermi velocity of electrons in graphene and is roughly of the order of $c/300$ ($c$ being the speed of light). Like S=1/2 spins
with spin-up and spin-down eigenstates, here the $\sigma$'s have two eigenvectors corresponding to two sublattices of the honeycomb lattice of graphene and thus
$\sigma$'s are called the pseudo-spins.

\begin{figure}
\includegraphics[width=.55\linewidth,angle=0]{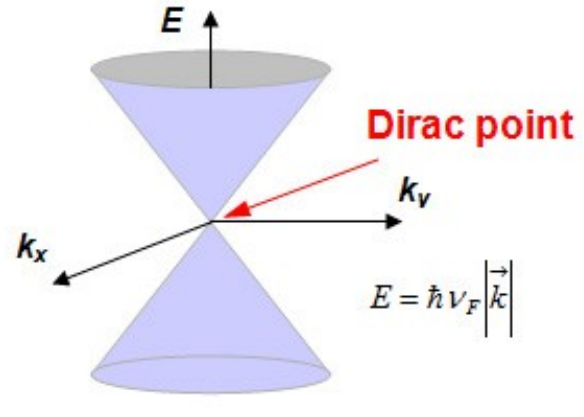}\hskip .3 in
\includegraphics[width=.29\linewidth,angle=0]{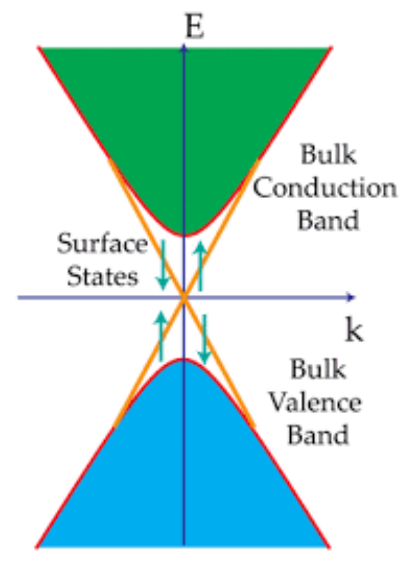}
\caption{(Left) Dirac point showing touching of conduction and valence bands in graphene $E-k$ diagram.
(Right) Topological insulator band structures with gapped bulk states and spin-filtered conducting edge/surface states.}
\label{fig1}
\end{figure}

\subsection*{Berry Curvature}
Electrons in a crystal moves around in presence of periodic ionic potentials.
In the nearly free electron model\cite{kittel}, electron feels
the ionic potential only when it is close to the periodic ionic positions and the wavefunctions $\psi_{{\bf k}}({\bf r})$ are described as 
\begin{eqnarray}
\psi_{{\bf k}}({\bf r})=e^{i{\bf k.r}}u_{{\bf k}}({\bf r}).
\end{eqnarray}
Apart from the free electron like factor $e^{i{\bf k.r}}$, the wavefunction $\psi_{{\bf k}}({\bf r})$ also
contains a {\bf Bloch function $u_{{\bf k}}({\bf r})$} that has the same periodicity
as the lattice.
Now {\bf for a time dependent Hamiltonian, the time evolution of such wavefunction accumulates an extra phase.} This is different from the usual dynamical phase that comes from a time-dependent Schrodinger equation.
Rather it is geometrical in nature for it depends only on the time dependence of the Bloch functions. This is called the Berry phase and for very slow
evolution of a system it can be shown to be equal to 
$\gamma=\int^{r(t)}<u_{{\bf k}}(r')|i\hbar\frac{\partial}{\partial r'}|u_{{\bf k}}(r')>dr'$ (when expressed 
in the parameter space $r(t)$), where $r'=r(t')$ denotes the electronic position at time $t'$.
In general, this integral depends on the path of temporal evolution for $r(t)$.
But this is path independent for a cyclic process, where the Hamiltonian is time periodic with $H(t)=H(t+T)$ and evolution 
only in steps of time periods $T$ are considered.
So {\bf for cyclic evolution $\gamma$ has a physical meaning and can be related to a physical variable.} However, the integrand $A_{{\bf k}}(r')=<u_{{\bf k}}(r')|i\hbar\frac{\partial}{\partial r'}|u_{{\bf k}}(r')>$, called a Berry connection, still possess a $k$-dependent phase degree of freedom that can be
inherently present in the Bloch functions.
 So Berry connection depends on the particular gauge chosen in the calculations\cite{shen}. 
\\
\fbox{%
\parbox{1.\linewidth}{%
A definite set of phase choice gives a definite gauge.}%
}
We can use Stoke's theorem to turn such closed line integral into a surface integral to get a {\bf gauge independent integrand $\Omega_{{\bf k}}(r)=\nabla\times A_{{\bf k}}(r)$, called Berry curvature. Due to the gauge independence
this can as well be related to a physical variable.}

 In the momentum space, the Berry phase for a cyclic evolution
can be defined (for $n$-th Bloch band $u_n(k)$\cite{kittel}) as a surface integral over the full BZ, of the Berry curvature $\Omega_n=\nabla_k\times <u_n(k)|i\nabla_k|u_n(k)>$. From there we get the {\bf Chern numbers\cite{shen} given as
$\nu=\sum_n\int_{BZ}\Omega_n d^dk$.
They stand for topological invariants 
of topological systems for such systems are characterized with
certain nonzero quantized values of these Chern numbers.} 
There are nontrivial systems where no single $u_n(k)$ function can be defined for the whole BZ and one need to consider multiple $u_n(k)$ functions, related 
to each other via mere phase differences, for different regimes of the BZ.
In those cases, the Berry phase or equivalently the Chern number takes nonzero quantized values and gives an quantitative estimate of the nontrivial topology of the system.

From electromagnetism, we know that magnetic field ${\bf B}$ and magnetic vector potential ${\bf A}$ are related by ${\bf B=\nabla\times A}$. ${\bf B}$ is a physically measurable quantity while ${\bf A}$ is not. Rather, ${\bf A}$ is gauge dependent and can always be replaced with ${\bf A+\nabla}\lambda~,\lambda$ being a scalar function. Hence ${\bf A}$ is called a gauge field.
In our present case,
 the gauge dependent Berry connection is
like a magnetic vector potential that varies depending on the particular gauge used. But its curl, $i.e.,$ the Berry curvature is gauge independent and is analogous to a magnetic field. In short, {\bf $\Omega_n$ is like a magnetic field in the momentum space.}

We find Berry curvature in graphene to identically become zero for all the Bloch vectors when both time reversal symmetry (TRS) and inversion symmetry (IS) is preserved 
[We should remind the reader here that a TRS implies $H(t)=H(-t)$ and an IS implies $H({\bf r})=H({\bf- r})$].
With inversion breaking, the Berry phase or the total surface integral still remains zero, even though nonzero contribution comes from the Dirac points K and K', around which line integral
of the Berry connection becomes $\pi$ and $-\pi$ respectively.

\subsection*{Topological Insulators}
TRS breaking comes with nonzero Chern numbers\cite{shen}. This amounts to topological nontriviality and causes conducting edge states in the system which would be impossible to get in the presence of TRS.\\
\fbox{%
\parbox{1.\linewidth}{%
Here~edge/surface~states~imply~the~states~localized~at~the~edges~of~an insulator. A topological insulator experiences conducting edge/surface states causing unidirectional electron/holes flow along the edges. This is a topological effect
characterized via nonzero Chern numbers.}}\\
A TI has a gapped energy spectrum ($i.e.,$ conduction and valence bands do not touch) making it an insulator in the bulk. However, their spectra becomes gapless at the edges allowing charge flow along the boundary.
For example in a quantum Hall effect (QHE), perpendicular magnetic field breaks TRS in a 2D electron gas to produce transverse
Hall conductivity: $\sigma_{\perp}=\nu\frac{e^2}{h},~\nu$ being the Chern number.
Later Haldane introduced unique magnetic flux distribution through the graphene/honeycomb lattice with zero overall flux through each hexagonal unit (amounting to an AC magnetic field). This adds mass terms ($i.e.,~m\sigma_z$ terms) of opposite signs in two valleys ($i.e.,$ K $\&$ K' points) in the graphene continuum model which ultimately results in nonzero quantized Chern numbers. In this case, the topological nontriviality appears without resorting to large DC magnetic fields, as required in QHE.
Such alternating flux distribution can also be achieved merely by adding a next nearest neighbor complex hopping term to
the graphene tight-binding model. Interestingly, one can also consider Kane-Mele's spinful model\cite{shen} where spin filtered
conducting edge states are obtained in presence of mirror symmetric spin-orbit coupling (SOC) for it produces two copies of Haldane model with opposite spins - popularly called quantum spin Hall effect (QSHE).\\
\fbox{%
\parbox{1.\linewidth}{%
SOC implies interaction a system experiences when spin and orbital
degrees of freedom couples~to($i.e.,$ interacts with) each~other.}}
\\
In fact, this is the famous toy model for describing a topological insulator (TI), where the bulk is insulating yet supporting conducting states at the edges and yet no TRS is broken. 
Just as a further note, we should point out that the topological edge states
of these systems remain intact even in the presence of the spin-z symmetry breaking additional Rashba-SOC term where the topological orders can be described via some $Z_2$ invariants\cite{shen}.
For more detailed discussion on TI, one can take a look at the Resonance article~\cite{reso}, written by one of the authors.

\subsection*{Floquet Topological Insulator}
At this point, it is not out of the way to discuss a bit of various possibilities and experimental realization of topological phases that the physics community is probing of late. The foremost of those will probably be the Floquet topological insulators (FTI). A trivial/non-topological quantum system can be driven out of equilibrium when a time periodic term is added to the system Hamiltonian. 
{\bf The idea of Floquet theory of time periodic systems is similar to the 
Bloch theory in space periodic lattices.}
As we look at the time periodic systems stroboscopically, $i.e.,$ in steps of
complete time periods, the wavefunctions can be written as products of
time periodic Floquet states and exponential factors.
This can lead to an effective stationary ($i.e.,$ time independent) Hamiltonian for the stroboscopic evolution of the originally time-periodic Hamiltonians\cite{moessner}.
Interestingly, such stationary Hamiltonian can behave topologically ($i.e.,~\nu\ne 0$) even though the original model is non-topological ($i.e.,~\nu=0$).
That's how we get a FTI. For example, an irradiation via circularly polarized light on graphene can create such systems\cite{moessner}. 
Just like we get nonzero quantized Chern numbers in a TI, we get similar
nonzero Chern numbers (called Floquet Chern number) for FTI as well.


\begin{figure}
\includegraphics[width=.6\linewidth,angle=0]{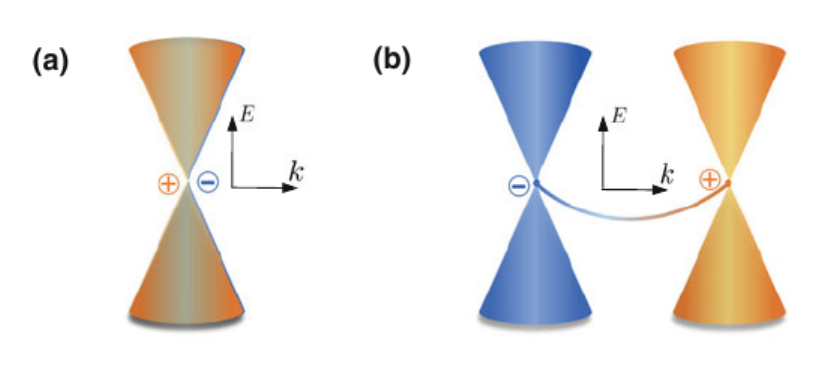}\\
\includegraphics[width=.6\linewidth,angle=0]{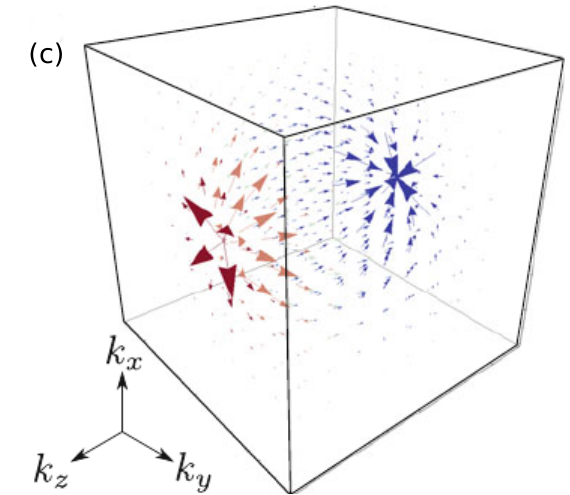}
\caption{(a) Dirac semimetal with positive and negative chiralities fermions sitting together. (b) Weyl semimetal where a DP get separated into two WP with opposite chiralities.
(c) distribution of Berry curvature ($i.e.,$ the field lines) in the 3D BZ in a Weyl semimetal. Figures are taken from Ref.\cite{shen}.}
\label{fig2}
\end{figure}
\section*{Weyl Semimetals}
All these models, so far mentioned are two-dimensional. However, topological effects are seen in higher dimensions as well. Of-course there are three dimensional (3D) TI materials, what we are mainly interested in the review are the ones called Weyl semimetals (WSM) ($e.g.,$ pyrochlore iridates). They can be looked upon as 3D analogue of graphene\cite{sumathi} though a WSM has both gapless surface and bulk states, unlike the graphene based topological insulators.
More specifically, the dispersion spectrum of the 3D bulk WSM system have 
discrete $k$-points, called Weyl points, where conduction and valence band touch
with each other linearly, as can be seen in Fig.\ref{fig2}. 
Besides that, there are gapless conducting surface states, localized only within the surface of the WSM.
We see that a graphene Hamiltonian does not have any $\sigma_z$ term due to IS.
Disregarding IS, if we add a $m\sigma_z$ term, graphene spectrum get gapped out. But in a similar 3D WSM bulk Hamiltonian $H=c_0(k)\sigma_0+\sigma_xv_xp_x+\sigma_yv_yp_y+\sigma_zv_zp_z$, such gaping out is not possible as degeneracies at Weyl nodes are accidental in nature and they are not outcome of the symmetries. This makes them more robust. One can only shift the position of those Weyl nodes but can not knock them out, unless pairs of nodes with opposite chiralities are made to coincide\cite{sumathi,aswin}.

While talking about chiralities one should again go back to see what that really means.
A three dimensional Dirac equation is represented via $4\times4$ matrices $\gamma^\mu$ ($\mu:~0,..,3$) that obey Clifford algebra (see Introduction section) among themselves.
The Dirac equation for a spin-$\frac{1}{2}$ massive particle is given as
\begin{eqnarray}
 (iv\hbar\gamma^\mu\partial_{\mu}-mv^2)\psi=0.
\end{eqnarray}
Here in our condensed matter context, we put velocity to be $v$, the Fermi velocity of the electrons, as opposed to $c$ that was used in Eq.\ref{dirac}. 
Also comparing with Eq.\ref{eq5}, we get the
the Hamiltonian density to be
$\mathcal{ H}=\psi^\dagger H\psi=-iv\hbar\bar{\psi}{\bf\gamma.\nabla}\psi+mv^2\bar{\psi}\psi$.\\
Here $\psi$ is a $4\times1$ complex spinor ($i.e.,$ a four-element column matrix with complex entries) and $\bar{\psi}=\psi^\dagger\gamma^0$.

Form of the $\gamma$ matrices depend on the representation considered. In the representation where $\gamma^0=\sigma_0\otimes\tau_x$
and $\gamma^i=\sigma_i\otimes i\tau_y$,
the Hamiltonian becomes
\begin{equation}
   H=
  \left[ {\begin{array}{cc}
  -v\sigma.p  & mv^2\sigma_0 \\
  mv^2\sigma_0  & v\sigma.p \\
  \end{array} } \right]
\label{hmatrix}
\end{equation}
where $p$ denotes the momentum operator.
Thus $\gamma$ matrices are the outer product of Pauli matrices $\tau$ and $\sigma$ (for spin and orbital subspaces respectively) given as,
\begin{equation}
   \sigma_0,\tau_0=
  \left[ {\begin{array}{cc}
   1 & 0 \\
  0  & 1 \\
  \end{array} } \right].~~~~~~~
  \sigma_1,\tau_x=
  \left[ {\begin{array}{cc}
   0 & 1 \\
  1  & 0 \\
    \end{array} } \right].\nonumber
\end{equation}
\begin{equation}
  \sigma_2,\tau_y=
  \left[ {\begin{array}{cc}
   0 & -i \\
  i  & 0 \\
  \end{array} } \right].~~~~~~~
  \sigma_3,\tau_z=
  \left[ {\begin{array}{cc}
   1 & 0 \\
  0  & -1 \\
  \end{array} } \right].
\label{hmatrix2}\nonumber
\end{equation}
For Weyl fermions, one has $m=0$ that makes $H$ block-diagonal
with $2\times2$ non-zero sub-blocks given by 
$\pm v\sigma.p$. It is under this massless condition, $H$ commutes
with the chirality operator 
\begin{equation}
 \boxed{\gamma_5=i~\Pi_{i=0}^d\gamma^i=i\gamma^0\gamma^1\gamma^2\gamma^3} ~~~~~\rm (in~ 3D)
\end{equation}
and the Weyl spinors become chiral.
In the representation considered, we obtain
$\gamma_5=-\sigma_0\otimes\tau_z$ (this would
become an identity matrix in even spatial dimensions\cite{aswin}).
In 3D, there are two chiral eigenstates, termed as right-chiral and left-chiral and
 the Hamiltonian is more conveniently expressed as $\chi v\sigma.p$ with chirality $\chi=\pm1$.
It signifies that Weyl fermions possess a definite chirality - either left or right. In this case, it simply means the direction of $\sigma$ and $p$ are either parallel or anti-parallel\cite{aswin}.

\subsection*{WSM spectrum within minimal models}
The form of the Hamiltonian \ref{hmatrix} (for $m=0$) refers to definite but opposite chirality Weyl fermion pairs. When they appear at same $k$-points,
they merge to become a Dirac point with overall zero chirality  (see the cartoon in Fig.\ref{fig2}(a)-(b)). 
Dirac points are protected by symmetry and can be undone by symmetry breaking.
for example in graphene, IS breaking causes
dispersion spectrum to be gapped thereby removing the Dirac points
 from the spectrum. But a WSM phase appears when time reversal breaking (TRB) or inversion breaking (IB) perturbations can not remove the gaplessness of the spectrum but only shift the positions of the
touching points. At that point, the Weyl points pairs, that made up the Dirac point, separate out to exist individually, each of the pair bearing opposite chiralities\cite{sumathi,aswin}. 
{\bf At those band touching points, the Berry curvatures are singular.
Thus the WP's can be interpreted as monopoles corresponding to the Berry curvatures
in the momentum space.} The charge of the Weyl node is given by the quantized Berry flux (a surface integral) around this point and is proportional to the Chern number (with an additional sign factor denoting the chirality).

Now consider $m$ is a varying parameter of the Weyl Hamiltonian (do not confuse $m$ with mass, which is already zero in Weyl fermion case)\cite{murakami}. Let the doubly degenerate DP's occur at $k=k_0$ for $m=m_0$ making the system a Dirac semimetal. This coincidence of the DP's is a result of symmetries and can be separated via breaking corresponding symmetries. We will look at the type of symmetries later. For now, let's consider that we get two separated gapless points (after separation, the DP becomes a WP pair) at, say, $k$ and $k'$ (close to $k_0$) when the varying parameter changes its value from $m_0$ to $m$.
Any two band (valence and conduction bands here) Hamiltonian, that is a function of both $k$ and $m$, can be written as 
\begin{eqnarray}
H(k,m)=a_0(k,m)\sigma_0+\sum_ia_i(k,m)\sigma_i.
\label{Hwsm}
\end{eqnarray}
 So the energy eigenvalues will be $E(k,m)=a_0(k,m)\pm\sqrt{\sum_ia_i(k,m)^2}$. Thus at $m=m_0$, $a_i(k_0,m_0)=0$ for the two bands to meet there.
Similarly, we should also have $a_i(k,m)=0$. The point 
$k$ being close to $k_0$, we can Taylor expand and get
$a_i(k,m)=\partial a_i/\partial k_j|_{k_0,m_0}(k_j-k_{0j})\sigma_i+\partial a_i/\partial m|_{k_0,m_0}(m-m_0)=0$.
Without loss of generality we can choose $m_0=0$ and
choose WSM phase to appear only for $m>0$.
So the nontrivial solution gives Det[$\partial a_i/\partial k_j|_{k_0,m_0}$]=0.
A thorough calculation\cite{murakami} shows that one need to go to at least one higher order in Taylor's expansion to get a meaningful result and accordingly  a typical Hamiltonian obtained in this way can be written as 
\begin{eqnarray}
\boxed{H(k,m)=\gamma(k_x^2-m)\sigma_1+v(k_y\sigma_2+k_z\sigma_3)}
\end{eqnarray}
\begin{figure}
\includegraphics[width=.9\linewidth,height=3.1 in,angle=0]{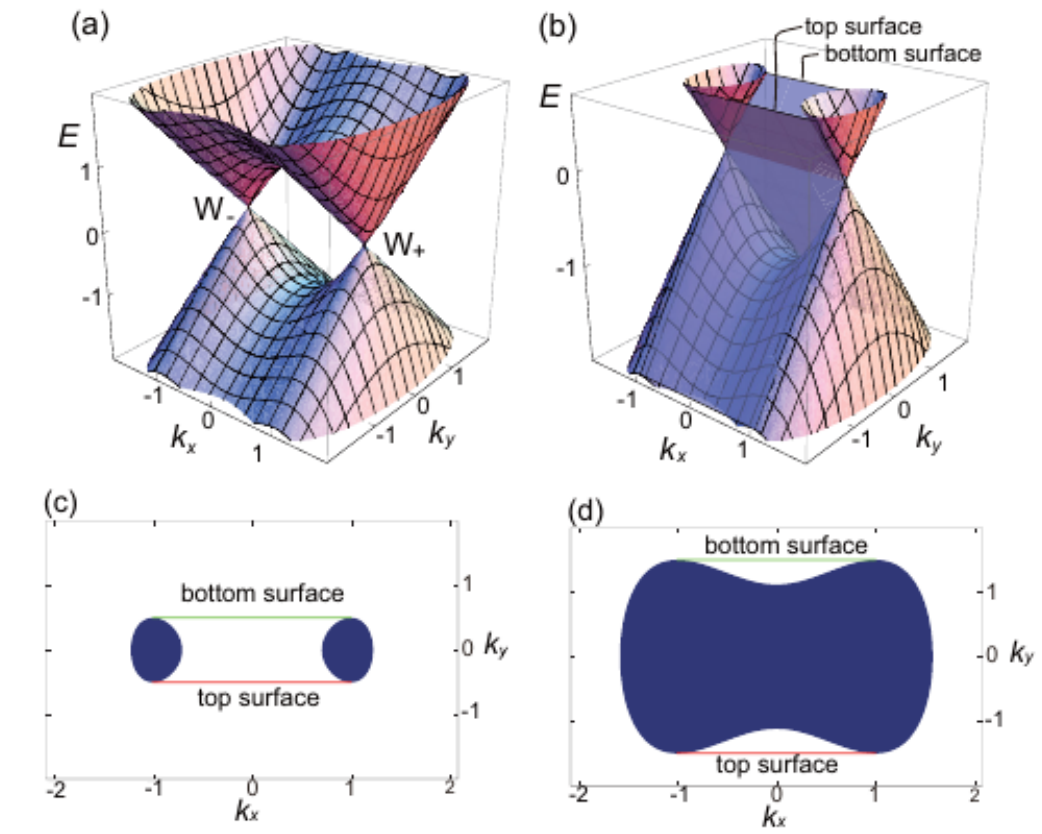}
\caption{Typical bulk (a,b) and surface (b) states of the model for WSM. Bulk Fermi surface and surface Fermi arcs are also shown for a typical (c) small and (d) large Fermi energy ($E_F$). Figures are taken from Ref.\cite{murakami}.}
\label{fig3}
\end{figure}

\subsection*{Bulk and surface states}
In the above Hamiltonian, bulk WSM phase appears for $0<m<k_{x,\rm max}^2$.
This is so as gapless Weyl points can occur only at $(\pm\sqrt{m},0,0)$ points
in the bulk spectrum.
A WSM phase breaks either TRS or IS or both (though in this particular case it breaks TRS alone), which need to be carefully found out (as this is the approximate continuum model and not the full lattice model\cite{trivedi}). Whenever two merged WP's of opposite chirality
get separated by some lack of symmetry, we get two WP's. Thus in a WSM, the number of WP's are even.
If TRS is broken, the minimum number of possible WP's is 2.
However, if IS is broken but TRS retained, there will be Kramer's partner for each $k$ vector. \\
\fbox{%
\parbox{1.\linewidth}{%
Kramer's theorem says that for a fermionic system with TRS, there always remains at least a two-fold degeneracy. Those two degenerate states, related via TRS,
are called Kramer's partner of each other.}}\\
So for each of the positions $k_1$ and $k_2$ where WP's are produced out of a doubly degenerate DP, there will be a Kramer's partner where another WP should be situated. Thus the minimum number of WP's become 4. We should also remember that a TRS broken, IS preserved WSM contain Weyl nodes at same energies while a further IS breaking disrupts the degeneracy of the nodes\cite{burkov}.

We see that WSM has Weyl points in an otherwise gapped bulk energy spectrum. 
Now if we consider a finite geometry of such materials, 
we find existence of surface states, $i.e.,$ states localized only at the boundary 
surface of the material (see Fig.\ref{fig3}).
For the model considered, it can be shown that surface states exists for $k_x^2<m$. They connect the conduction and valence bands. So they are gapless. Moreover, the surface states feature Fermi arcs (Fig.\ref{fig3}(c)-(d)) joining the bulk Fermi pockets (for low energy) arising out of two WP of opposite chiralities\cite{murakami}.
\\
\fbox{%
\parbox{1.\linewidth}{%
In general, Fermi arc means open, unclosed contour in $k$-space where the dispersion energy is same as the Fermi energy. In our case, if we imagine a constant energy cut (consider that to be the Fermi energy) within the dispersion spectra, the bulk-dispersion cone coming out of a WP will produce a closed loop. These are called Fermi pockets. Additionally, that energy cut in the surface-dispersion spectra gives line segments joining those Fermi pockets. These lines are the Fermi arcs. See Fig.\ref{fig3}.}}\\

\subsection*{Type I and Type II WSM}
\begin{figure}
\includegraphics[width=\linewidth,angle=0]{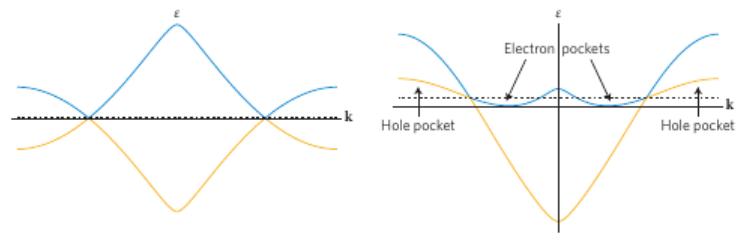}
\caption{Type I and type II Weyl semimetals. Figures are taken from Ref.\cite{chiral}.}
\label{type}
\end{figure}

We should mention here that the $k$-dependent term $a_0(k,m)$ in the Hamiltonian \ref{Hwsm} can add interesting modification to the Weyl bulk spectrum. In absence of any $k$ dependence there, the density of state (DOS) at the node energy, is vanishingly small and the Fermi surfaces are just discrete Weyl points. But an adequate $k$ dependence induces a crystal field anisotropy in the band
dispersion near a WP\cite{aswin} and can tilt the Weyl cones creating electron and hole pockets at the node energy. The cartoon in Fig. \ref{type} illustrates such different situations. In the former case with zero DOS at the Weyl node, we get what is called a type I WSM while in the latter case, that supports Fermi pockets at node energy, is named as type II WSM. One phase transits to the other one via a transition called Lifshitz transition\cite{chiral}.

Layered transition-metal dichalcogenide $WTe_2$ was the first predicted type-II WSM. But as the surface Fermi arcs there were short in length, another compound $MoTe_2$ was proposed later where Fermi arc length is much longer and thus easily detectable in angle-resolved photo-emission spectroscopy (ARPES)\cite{yan}.
Also in $MoTe_2$, Weyl nodes situate much closer to Fermi energy and thus Weyl physics are observed as low energies.

\subsection*{Why Topological ?}
At this stage, we may wonder what is topology to do with this topic and why these WSM phases are called topological.
Naively speaking, topology describes the state of an object that can not change via any continuous transformation. For example, a circle and a square can be deformed to each other and that's why they fall within same topological class. However, a donut with a hole within represent a different topology. We define a quantity called ``topological invariant'' that remains the same in one particular topology. We already discussed that conducting edge states in a 2D TI are topological and the corresponding topological invariants are called Chern numbers. To change the currents corresponding to those edge states, we need to change the topology by altering the Chern number, which is a quantized number (only integers, in this case). Hence those edge states are very robust against small impurities or doping. Similarly, the band-touching Weyl points in a 3D WSM bulk energy spectrum present a topologically robust feature. Unlike in the 2D case,
here we compute a closed surface integral of the Berry curvature surrounding
a Weyl point. This is called the chiral charge for in 3D, the Berry flux ($i.e.,$ surface integral of the Berry curvature) acts like
a magnetic field\cite{aswin} (see Fig. \ref{fig2}c). It also stands for the topological invariant in the system. Thus each Weyl point carry with themselves a topologically
invariant chiral charge and in case when two Weyl points of opposite chirality
merge with each other, it gives a Dirac point whose overall chirality is zero.
So the value of topological invariant becomes zero as well (making the system topologically trivial). 
Therefore, within the same topological class, the opposite-chirality WP pair can only be shifted (via tuning of system parameters) but not removed 
from the spectrum.

WSM also feature topological surface states characterized with Fermi arcs. 
Fermi arcs are open lines in $k$-space
that connects surface-projected Weyl points with opposite chirality\cite{yang}. Fermi arcs are not like usual Fermi surfaces 
that are closed surfaces in the momentum space. 
In a WSM, the Chern numbers are proportional to the chiral charges
of the Weyl nodes. They are topologically protected and hence conserved
in each WP. However, interesting modification comes in presence of an electric (E) and magnetic (B) field. This is called chiral anomaly. We will discuss this briefly
in the following section.

\subsection*{Presence of electromagnetic field - Chiral anomaly}
By Noether's theorem\cite{noether}, there is always a conserved current (consider a four-vector $J^\mu$) associated with a continuous global symmetry.
Writing it out mathematically, a current conservation, in the four-vector language, implies $\partial_\mu J^\mu=0$.
This leads to the equation of continuity and thereby the conservation of charge $\int  J^0d^3x$.

Now the axial vector current of the Dirac theory is given by, $J_5^\mu=\bar\psi\gamma^\mu\gamma_5\psi$.
So in a WSM With chiral symmetry, we should have
$\partial_\mu J_5^\mu=0$ and consequently, the conservation of
chiral charge at the Weyl nodes\cite{chiral}. This charge, apart from a prefactor,
is the Chern number $C$ at that WP in the WSM. This can be shown to be same as a closed surface integral of Berry curvature in $k$-space around the WP.
That's why Weyl nodes act like magnetic monopoles corresponding to 
Berry curvature in the momentum space. The non-zero chiral charge is an outcome
of the band touching at WP and singularity of the Berry curvature there.

However a coupling to external gauge field (coming from electromagnetic fields) changes the situation.
It invokes canonical momentum
to the problem modifying the Weyl equation to be $i\gamma^\mu(\partial_\mu+iA_\mu)\psi=0$ where $A^\mu$ is the electromagnetic four potential.
This modifies the axial current conservation relation to be $\partial_\mu J_5^\mu=\frac{1}{8\pi^2}F^{\mu\nu}F_{\mu\nu}$ with $F_{\mu\nu}=(\partial_\mu A_\nu-\partial_\nu A_\mu)$.
Written in terms of the electric (E) and magnetic (B) field, this gives
\begin{equation}
\boxed{\partial_\mu J_5^\mu=\frac{1}{4\pi^2}{\bf E.B}}
\end{equation}
in 3D.
This is the essence of chiral anomaly which points to non-conservation of chiral charge at individual Weyl nodes (producing nodal/valley polarizations with source and sink of charges\cite{aswin}) and results in chiral flow between the nodes of opposite chiralities, when external electric and magnetic field, not perpendicular to each other, remains present.

In 1 spatial dimension, there is effectively no magnetic field\cite{princeton} and electric field alone creates the chiral anomaly. But in 3D, electrodynamics allows the presence of both ${\bf E~\&~B}$\cite{princeton}. Magnetic field creates Landau levels that disperse only along ${\bf B}$ direction (and degenerate along directions normal to it).
However, only the zeroth Landau level is chiral with charge propagation along/opposite to the field direction and velocities being reversed for opposite chirality Weyl nodes\cite{prl119}. For this, the 3D problem is effectively like a 1D problem with electrons propagating along the magnetic field lines forming chiral 1D channels.

\section*{Summary and Discussions}
Let us now summarize as well as highlight the utility of studying and finding the Weyl semimetals.
What makes them special?
The first thing to note that this is {\bf a major discovery in the hunt for Weyl fermions.} WSM are the condensed matter systems to first show the existence of Weyl fermions\cite{yan}.
For long, Neutrinos were thought of as possible candidate for Weyl fermions until recent discovery that found Neutrinos also to possess some mass (and hence disqualified to be a Weyl fermion).
Thus the recent discoveries of WSM materials had a huge and deep impact in the science community.
In a WSM, we find that the Berry curvature of the WSM Bloch bands, which is
like a magnetic field in momentum space, produces topologically robust magnetic monopoles due to the presence of Weyl nodes in the bulk energy spectrum.
Furthermore, in presence of non-orthogonal electric and magnetic field, a WSM
exhibits chiral anomaly which in turn can give rise to negative magnetoresistance\cite{yan}. All these exotic behaviors make a WSM very special in the field of quantum condensed matter leading to rapid-fire researches in recent times.

Now the discussion remains as to which are the WSM materials
and what restrictions we need to provide to those materials in order to
 witness the Weyl physics there. It turns out that one needs to find WSM, with all Weyl nodes 
related by symmetry and close to the Fermi energy $E_F$.
They also need to be far apart in the $k$-space and with no non-topological bands close by so that the Weyl nodes can be easily singled out for probing and experimenting\cite{aswin}.
In 2015, the compounds $TaAs,~TaP,~ NbAs$ and $NbP$ were found to show WSM behavior with Fermi arcs observed in Angle Resolved Photo-emission Spectroscopy (ARPES) measurements. 
Pyrochlore Iridates like $Y_2Ir_2O_7$ or $HgCr_2Se_4$ have also been predicted as WSM materials\cite{yan}.

 Lastly, we want to add that a WSM is highly mobile due to its gapless spectrum.
 They possess topologically protected Weyl points, surface Fermi arcs and are chiral possessing spin-momentum locking with spins aligned/ anti-aligned with the momentum directions. These days, Weyl semimetals are turning out to be a rapidly evolving field of study in condensed matter as a WSM can be used in a plethora of spintronic, chiral or valleytronic applications\cite{sumathi,aswin}.
 
 \section*{Acknowledgements}
 Both SK and AMJ thank DST, India for financial support (through
Start-Up research grant and J. C. Bose National Fellowship respectively).


\end{document}